\RequirePackage[mathlines]{lineno} % Display line numbers
\documentclass[a4paper,11pt]{article}
\pdfoutput=1 % if your are submitting a pdflatex (i.e. if you have
\usepackage{jheppub} % for details on the use of the package, please
\usepackage[T1]{fontenc} % if needed

\usepackage{threeparttable}%This package is used for tablenotes
\usepackage{multirow}
\usepackage{subfigure}
\usepackage{float}
\let\oldequation\equation
\let\oldendequation\endequation
\renewenvironment{equation}
  {\linenomathNonumbers\oldequation}
  {\oldendequation\endlinenomath}
\def \jpsi {J/\psi}
\def \denu  {D^-e^+\nu_e}
\def \ee   {e^+e^-}

\def \gev  {\mbox{GeV}}

\def \gevcc{\mbox{GeV/$c^2$}}
\def \mev  {\mbox{MeV}}
\def \mevc  {\mbox{MeV/$c$}}

\begin{document}
\setrunninglinenumbers
%\begin{linenumbers}

\title{\boldmath Search for the rare semi-leptonic decay $\jpsi\to\denu+c.c.$}

\collaborationImg{\includegraphics[height=30mm,angle=90]{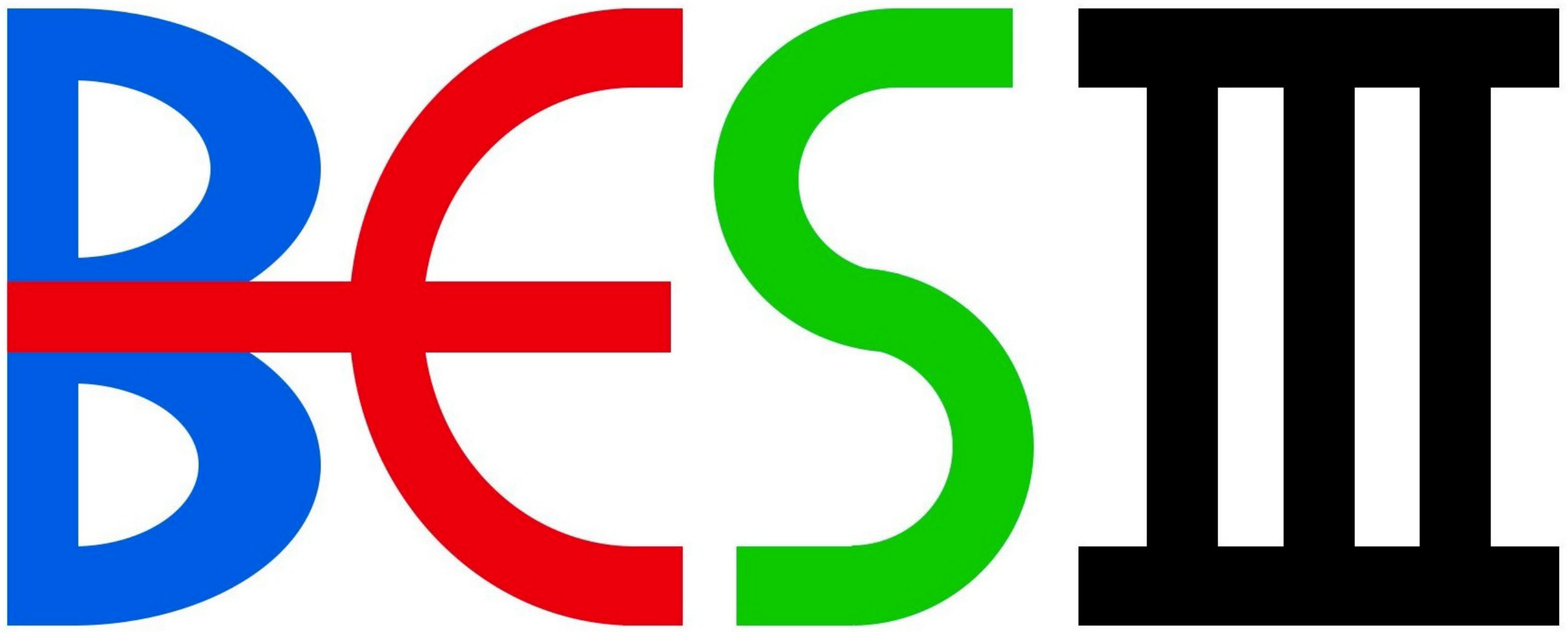}}
\collaboration{The BESIII collaboration}
\emailAdd{besiii-publications@ihep.ac.cn}

\abstract{
Using $10.1\times10^{9}$ $\jpsi$ events produced by the Beijing Electron Positron Collider~(BEPCII) at a center-of-mass energy $\sqrt{s}=3.097~\rm{GeV}$ and collected with the BESIII detector, we present a search for the rare semi-leptonic decay $\jpsi\to\denu+c.c$.
No excess of signal above background is observed, and an upper limit on the branching fraction $\mathcal{B}(\jpsi\to D^{-}e^{+}\nu_{e}+c.c.)<7.1\times10^{-8}$ is obtained at $90\%$ confidence level.
This is an improvement of more than two orders of magnitude over the previous best limit.
}

\keywords{$\ee$ experiments, charmonium, rare semi-leptonic weak decay}

\arxivnumber{2104.06628}

\maketitle
\flushbottom

\section{INTRODUCTION}
\label{sec:introduction}
\hspace{1.5em}

The $\jpsi$ meson, which decays primarily through strong and electromagnetic interactions, has been thoroughly studied for decades. However, its weak decays remain elusive.
Since the $\jpsi$ mass is below the $D\bar{D}$ threshold, the $\jpsi$ resonance is forbidden to decay into a pair of charmed mesons. 
However, it can decay into a single charmed meson accompanied by light hadrons or leptons via weak decay of one of the charm quarks.
The inclusive branching fraction (BF) of weak decays to a single charmed meson was predicted to be at the order of $10^{-8}$ or below~\cite{verma:1990, Sanchis:1994, Sanchis:1993, sharma:1999, wang:2008a, wang:2008b, shen:2008, dhir:2013, ivanov:2015, tian:2017} in the Standard Model (SM). Therefore, searching for these decays not only tests the SM prediction~\cite{Ablikim:2019hff}, but also probes new physics theories beyond the SM, such as the Top-color model, the Minimal Supersymmetric SM with or without R-parity, and the two-Higgs doublet model~\cite{zhang:2001, li:2012, datta:1999, hill:1995}, in which these BFs could be significantly larger, reaching values of $10^{-5}$~\cite{tian:2017}. 
So far, weak decays of the $\jpsi$ meson have not yet been observed~\cite{pdg:2020, bes:2006, bes:2008, bes3:2014, bes3:2017}. 

\begin{figure}[htbp] \centering
	\setlength{\abovecaptionskip}{-1pt}
	\setlength{\belowcaptionskip}{10pt}
	\includegraphics[width=10.0cm]{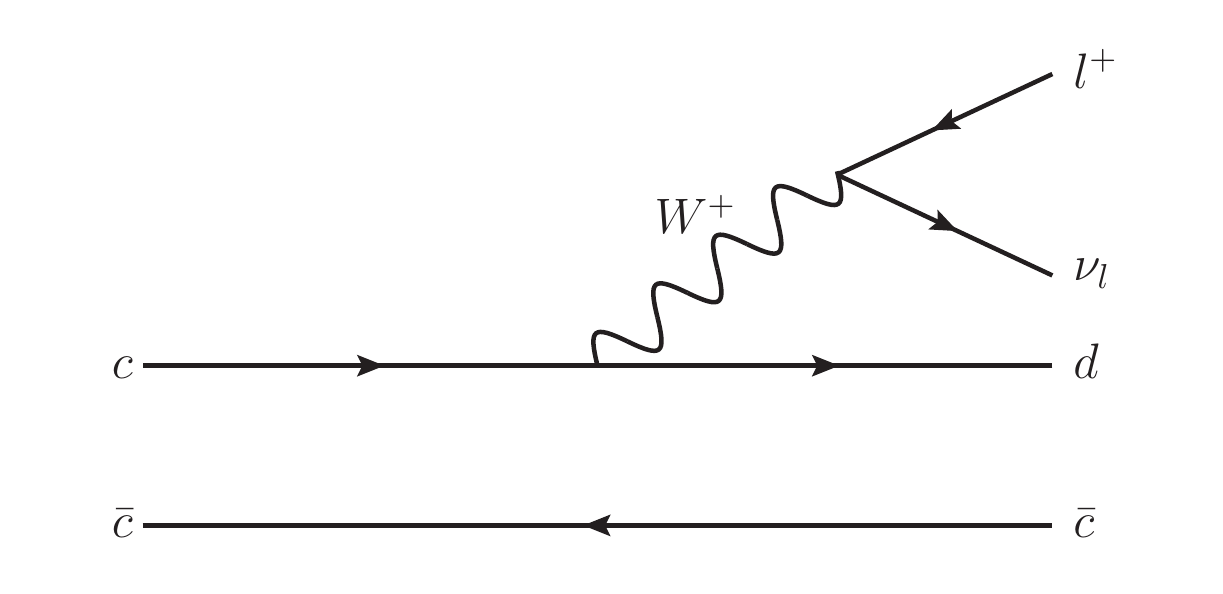}
	\caption{Feynman diagram for $\jpsi\to D^{-}l^{+}\nu_{l}$ decays at tree-level.}
	\label{fig:feynman}
\end{figure}

%Among the $J/\psi$ weak decays, the weak semi-leptonic decays suffer from minimal effects 
%from the strong interactions and their theoretical calculations have better controllable uncertainties, 
In weak semi-leptonic $J/\psi$ decays, the hadronic transition form factor between the initial and final-state mesons can be cleanly decoupled from the weak current~\cite{wang:2008b, shen:2008, dhir:2013, ivanov:2015, tian:2017}. 
Figure~\ref{fig:feynman}~shows the tree-level Feynman diagram within the SM for the decays $\jpsi\to D^{-}l^{+}\nu_{l}$~($l=e$ or $\mu$).
The theoretical predictions for the BF of the rare semi-leptonic decay $\jpsi\to D^{-}e^{+}\nu_{e}$ within the SM are of the order of $10^{-11}$~\cite{wang:2008b, shen:2008, dhir:2013, ivanov:2015, tian:2017}, as shown in Table~\ref{tab:prediction}.
A previous study of this decay by the BES collaboration reported an upper limit (UL) on the BF of $1.2\times10^{-5}$ at $90\%$ confidence level (CL) based on a sample of $5.8\times10^{7}~\jpsi$ events~\cite{bes:2006}. This result reaches down to the level of the expected BF values in several models beyond the SM~\cite{datta:1999, hill:1995}, although it is several orders of magnitude larger than the SM value. To further test the SM predictions and constrain the contributions from new physics models, a new measurement of $\mathcal{B}(\jpsi\to D^{-}e^{+}\nu_{e})$ with greater sensitivity is required.

\vspace{-0.0cm}
\vspace{-0.0cm}

\begin{table*}[!htbp]
\caption{Theoretical results for the BF of the semi-leptonic decay $\jpsi\to D^{-}e^{+}\nu_{e}$ ($\times10^{-11}$).}
\setlength{\abovecaptionskip}{1.2cm}
\setlength{\belowcaptionskip}{0.2cm}
\begin{center}
\footnotesize
\vspace{-0.0cm}
\begin{tabular}{l|cccccccccc}
\hline \hline
                    Decay mode & QCDSR~\cite{wang:2008b} & LFQM~\cite{shen:2008} & BSW~\cite{dhir:2013} & CCQM~\cite{ivanov:2015} & BSM~\cite{tian:2017}\\
                    \hline
                    $\jpsi\to D^{-}e^{+}\nu_{e}$ & $0.73_{-0.22}^{+0.43}$ & 5.1-5.7 & $6.0_{-0.7}^{+0.8}$ & 1.71 & $2.03_{-0.25}^{+0.29}$\\
\hline \hline
\end{tabular}
\label{tab:prediction}
\end{center}
\end{table*}
\vspace{-0.0cm}

In this paper, we report a search for the semi-leptonic decay $\jpsi\to D^{-}e^{+}\nu_{e}+c.c.$ with $D^{\pm}\to K^{\mp}\pi^{\pm}\pi^{\pm}$ using 
$10.1\times10^{9}$ $\jpsi$
events collected at the center-of-mass energy $\sqrt{s}=3.097~\rm{GeV}$ with the BESIII detector~\cite{Ablikim:2009aa} operating at the Beijing Electron Positron Collider (BEPCII)~\cite{Yu:IPAC2016-TUYA01}.
In order to avoid possible bias, we first validate the analysis with about $10\%$ of the full data sample. The final result is obtained with the full data sample by repeating the validated analysis strategy.
In addition, Monte Carlo (MC) simulation samples are used to optimize the event selection criteria, determine the signal detection efficiency and study the background. Throughout this paper, the charge-conjugate processes are always implied.

\section{BESIII DETECTOR AND MONTE CARLO SIMULATION}
\label{sec:detector}
\hspace{1.5em}
The BESIII detector~\cite{Ablikim:2009aa} records symmetric $e^+e^-$ collisions 
provided by the BEPCII storage ring~\cite{Yu:IPAC2016-TUYA01}, which operates with a peak luminosity of $1\times10^{33}$~cm$^{-2}$s$^{-1}$
in the center-of-mass energy range from 2.0 to 4.95~GeV.
BESIII has collected large data samples in this energy region~\cite{Ablikim:2019hff}. The cylindrical core of the BESIII detector covers 93\% of the full solid angle and consists of a helium-based
 multilayer drift chamber~(MDC), a plastic scintillator time-of-flight
system~(TOF), and a CsI(Tl) electromagnetic calorimeter~(EMC),
which are all enclosed in a superconducting solenoidal magnet
providing a 1.0~T (0.9~T in
2012) magnetic field. The solenoid is supported by an
octagonal flux-return yoke with resistive plate counter muon
identification modules interleaved with steel. 

The charged-particle momentum resolution at $1~{\rm GeV}/c$ is
$0.5\%$, and the $dE/dx$ resolution is $6\%$ for electrons
from Bhabha scattering. The EMC measures photon energies with a
resolution of $2.5\%$ ($5\%$) at $1$~GeV in the barrel (end cap)
region. The time resolution in the TOF barrel region is 68~ps, while
that in the end cap region is 110~ps. The end cap TOF
system was upgraded in 2015 using multi-gap resistive plate chamber
technology, providing a time resolution of
60~ps~\cite{etof1, etof2}.

%Simulated samples, produced with the {\sc geant4}-based~\cite{geant4} MC simulation software BOOST~\cite{bes:boost} which includes the geometric and material description~\cite{geo1,geo2} of the BESIII detector, the detector response and digitization models, are used to determine the detection efficiency and estimate physical backgrounds.
%The analysis is performed under the BESIII offline software system (BOSS)~\cite{bes3:boss705} which incorporates the detector calibration, event reconstruction and data storage.
%The simulation takes the beam-energy spread and initial-state radiation (ISR) in the $\ee$ annihilations into account, modeled with the {\sc kkmc}~\cite{ref:kkmc} generator.
%The inclusive MC sample consists of the production of the $\jpsi$ resonance and the continuum processes incorporated in {\sc kkmc}~\cite{ref:kkmc}.
%By assuming the decay $\jpsi\to D^{-}e^{+}\nu_{e}$ is dominated by the weak interaction via $c\to d$ charged current process and simply ignoring the hadronization effects and quark spin-flip, the signal MC events are generated in {\sc evtgen}~\cite{ref:evtgen}.
%The known $\jpsi$ decay modes are modeled with {\sc evtgen}~\cite{ref:evtgen} using BFs taken from the Particle Data Group~\cite{pdg:2020}, and the remaining unknown decays are generated by {\sc lundcharm}~\cite{ref:lundcharm}. 

Simulated data samples produced with the {\sc
geant4}-based~\cite{geant4} MC package BOOST~\cite{bes:boost}, which
includes the geometric and material description of the BESIII detector~\cite{geo1,geo2} and the
detector response, are used to determine detection efficiencies
and to estimate backgrounds. The simulation models the beam
energy spread and initial state radiation (ISR) in the $e^+e^-$
annihilations with the generator {\sc
kkmc}~\cite{ref:kkmc1, ref:kkmc2}. 
The inclusive MC sample includes both the production of the $J/\psi$
resonance and the continuum processes incorporated in {\sc
kkmc}~\cite{ref:kkmc1, ref:kkmc2}.
By assuming the decay $\jpsi\to D^{-}e^{+}\nu_{e}$ is governed by the weak interaction via a $c\to d$ charged current process, and ignoring the hadronization effects and quark spin-flip~\cite{bes3:2014}, signal MC events are generated in {\sc evtgen}~\cite{ref:evtgen1, ref:evtgen2}.
The known $\jpsi$ decay modes are modelled with {\sc
evtgen}~\cite{ref:evtgen1, ref:evtgen2} using BFs taken from the
Particle Data Group~\cite{pdg:2020}, and the remaining unknown charmonium decays are modelled with {\sc lundcharm}~\cite{ref:lundcharm1, ref:lundcharm2}. Final state radiation~(FSR)
from charged final state particles is incorporated using the {\sc
photos} package~\cite{photos}.

\section{EVENT SELECTION AND DATA ANALYSIS}
\label{sec:analysis}
\hspace{1.5em} 
The analysis is performed with the BESIII offline software system (BOSS)~\cite{bes3:boss705} which incorporates the detector calibration, event reconstruction and data storage. In the signal process $\jpsi\to D^{-}e^{+}\nu_{e}$, $D^{-}\to K^{+}\pi^{-}\pi^{-}$, 
we detect all final-state particles except the $\nu_{e}$. Charged tracks detected in the MDC are required to be within a 
polar angle ($\theta$) range of $|\rm{cos\theta}|<0.93$, where 
$\theta$ is defined with respect to the $z$-axis.
%All charged tracks must satisfy a detection acceptance of $|\rm{cos}\theta|<0.93$, where $\theta$ is the polar angle with respect to the beam axis.
Selected charged tracks are required to satisfy $R_{xy}<1.0$~cm and $|V_{z}|<10$~cm, where $R_{xy}$ and $|V_{z}|$ are the distances of closest approach to the interaction point of the track in the plane perpendicular to the beam and along the beam direction, respectively.
We retain the events with exactly four selected charged tracks with zero net charge. 
Particle identification~(PID) for charged tracks combines measurements of the energy deposited in the MDC~(d$E$/d$x$) and the flight time in the TOF to form likelihoods $\mathcal{L}(h)~(h=p,K,\pi)$ for each hadron $h$ hypothesis.
The charged kaons and pions are identified by comparing the likelihoods for the kaon and pion hypotheses, $\mathcal{L}(K)>\mathcal{L}(\pi)$ and $\mathcal{L}(\pi)>\mathcal{L}(K)$, respectively.
Positron PID uses the measured information in the MDC, TOF and EMC. The combined likelihoods ($\mathcal{L}'$) under the positron, pion, and kaon hypotheses are obtained.
Positron candidates are required to satisfy $P_e>0.001$ and $P_e/(P_\pi+P_K)>4$, while $\pi$ ($K$) candidates fulfil the criteria $P_\pi>P_K$ ($P_K>P_\pi$).
To further reduce background from hadrons, the ratio of the deposited energy of the positron candidate in the EMC, $E$, and its momentum obtained in the MDC, $p$, is required to be in the range $0.85<E/p<1.05$.

Neutral showers deposited in the EMC crystals are identified as photon candidates when the shower energies are larger than 25~MeV in the barrel ($|\cos{\theta}|<0.8$) and 50~MeV in the end cap ($0.86<|\cos{\theta}|<0.92$). 
In order to suppress fake photons due to electronic noise or beam background, the shower clusters are required to be detected within [0, 700]~ns from the event start time. 
In addition, photon candidates must be at least $10^{\circ}$ away from any charged tracks to remove fake photons caused by hadronic showers or final state radiations.

The selected charged hadron candidates, $K^+\pi^-\pi^-$, are used to form the $D^{-}$ meson.
Its invariant mass $M_{K\pi\pi}$ is required to be within the range of $[1.85, 1.89]~\gevcc$, corresponding to $\pm3$ times the mass resolution around the $D^-$ known mass~\cite{pdg:2020}.
A kinematic fit constraining the $K^+\pi^-\pi^-$ invariant mass to the $D^-$ mass~\cite{pdg:2020} is performed and the fit $\chi^2_{\rm{1C}}$ value is required to be less than 10.
To suppress background contributions from mis-identified events with extra photons, we require the total energy of good photons ($E^{\rm{tot}}_{\gamma}$) to be less than 0.2~$\gev$.

Due to conservation of energy and momentum, the undetected neutrino $\nu_e$ carries a missing-energy $E_{\rm{miss}}=E_{\jpsi}-E_{D^-}-E_{e^+}$ and a missing-momentum $\vec{p}_{\rm{miss}}=\vec{p}_{\jpsi}-\vec{p}_{D^-}-\vec{p}_{e^+}$, 
where $E_{D^-}$ ($E_{e^+}$) and $\vec{p}_{D^-}$ ($\vec{p}_{e^+}$) are the energy and momentum of the $D^-$ ($e^+$) in the rest frame of the initial $e^+e^-$ collision.
In order to suppress the background contributions from $\jpsi$ hadronic decays in which a pion or a kaon is mis-identified as a positron, $|\vec{p}_{\rm{miss}}|$ is required to be larger than 50~$\mevc$.
We extract the yield of the signal decays by examining the variable $U_{\rm{miss}}=E_{\rm{miss}}-c|\vec{p}_{\rm{miss}}|$, in which the signal candidates are expected to peak around zero if the final states of the semi-leptonic decay have been identified correctly.

\begin{figure}[htbp] \centering
	\setlength{\abovecaptionskip}{-1pt}
	\setlength{\belowcaptionskip}{10pt}
	\includegraphics[width=10.0cm]{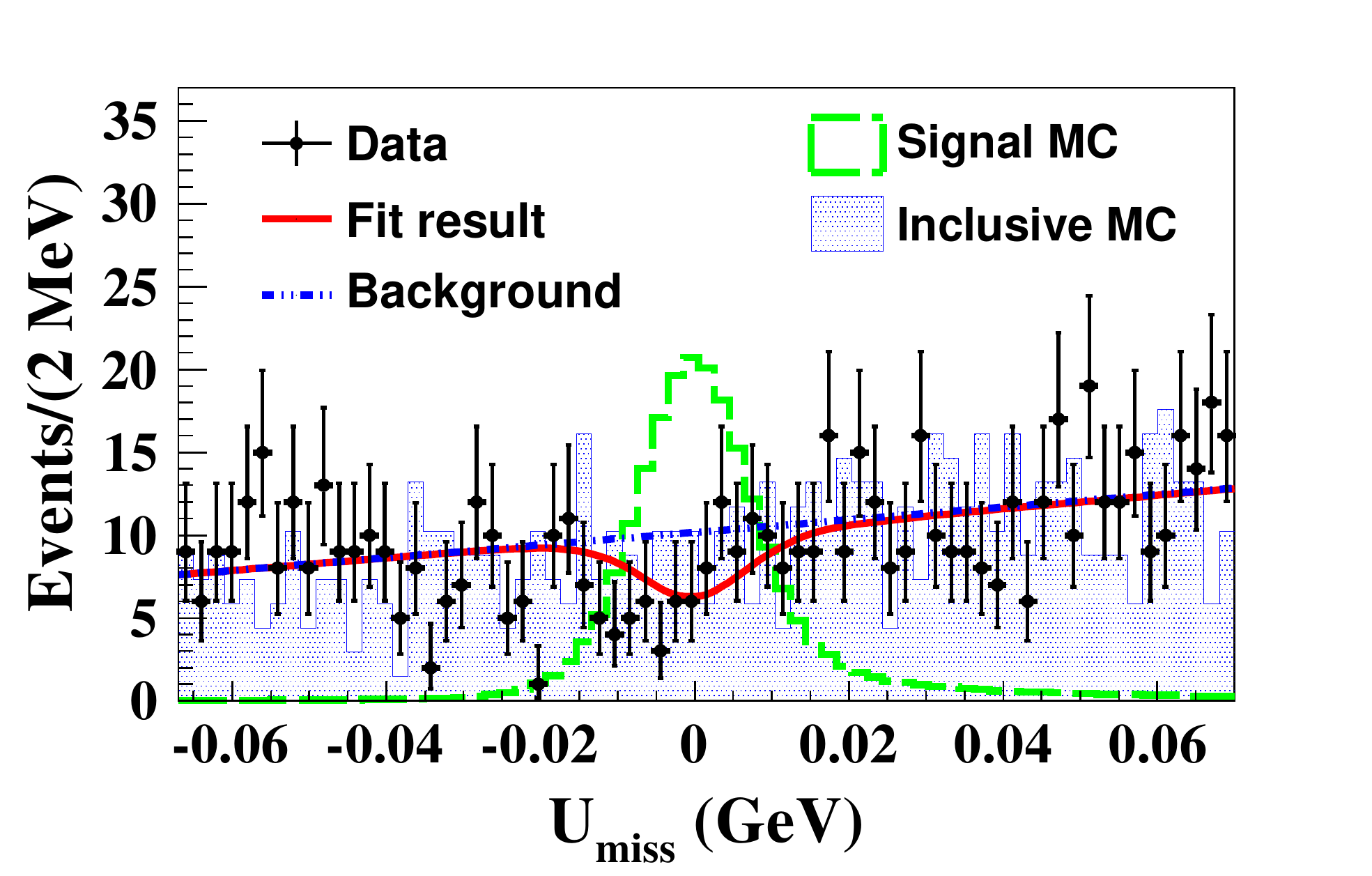}
	\caption{The $U_{\rm{miss}}$ distributions and the unbinned maximum likelihood fit. The black dots with error bars are data, the red solid line is the total fit result, and the blue dotted-dashed line is the background. The green long-dashed histogram shows the signal MC simulation events and the blue shaded histogram represents the inclusive MC events. Here, the signal MC events histogram is drawn with an arbitrary normalization, while the inclusive MC events histogram and the fit curve are normalized to the data luminosity.}
	\label{fig:umiss_fit}
\end{figure}

Figure~\ref{fig:umiss_fit} shows the  $U_{\rm{miss}}$ distribution in data, where no clear enhancement around zero is observed.
Using signal MC simulation,
%With all requirements mentioned previously, based on simulated $\jpsi\to D^{-}e^{+}\nu_{e}$ signal events,
the detection efficiency for $\jpsi\to D^{-}e^{+}\nu_{e}$ passing all selection requirements is determined to be $(29.93\pm0.10)\%$, where the uncertainty is statistical. The background contributions are investigated using an inclusive MC simulation sample, whose size corresponds to that in data~\cite{ref:topoana}.
As shown in Figure~\ref{fig:umiss_fit}, the $U_{\rm{miss}}$ distribution in the inclusive MC simulation sample is consistent with that in data and no peaking structure is seen around the signal region. %Since no significant signal is observed, a UL on $\mathcal{B}(\jpsi\to\denu)$ is estimated.<--- too early

\vspace{-0.0cm}
\vspace{-0.0cm}

\section{RESULT}
\label{sec:result}
\hspace{1.5em}
An unbinned extended maximum likelihood fit is used to estimate the signal yield.
The probability density function of the signal is derived from the shape of signal MC simulation of the $U_{\rm{miss}}$ spectrum, while the background shape is modeled with a linear function. 
%The fit result is shown in Figure~\ref{fig:umiss_fit}. 
As shown in Figure~\ref{fig:umiss_fit}, a negative signal is obtained, which indicates no signal is found from the fit result.
The BF of the signal decay is calculated as
\begin{eqnarray}
\mathcal{B}(\jpsi\to D^{-}e^{+}\nu_{e}+c.c.)=\frac{N_{\rm{signal}}}{N_{\jpsi}\times\epsilon\times\mathcal{B}_{\rm{sub}}},
\end{eqnarray}
where $N_{\rm{signal}}$ is the number of signal decays, $N_{\jpsi}=(10087\pm44)\times10^{6}$ is the number of $\jpsi$ events determined with the method described in Ref.~\cite{bes3:njpsi2012}, $\epsilon$ is the signal detection efficiency, and $\mathcal{B}_{\rm{sub}}$ is the BF of the intermediate decay $D^{\pm}\to K^{\mp}\pi^{\pm}\pi^{\pm}$ quoted from Ref.~\cite{pdg:2020}.

\begin{figure}[tp] \centering
	\setlength{\abovecaptionskip}{-1pt}
	\setlength{\belowcaptionskip}{10pt}
	\includegraphics[width=10.0cm]{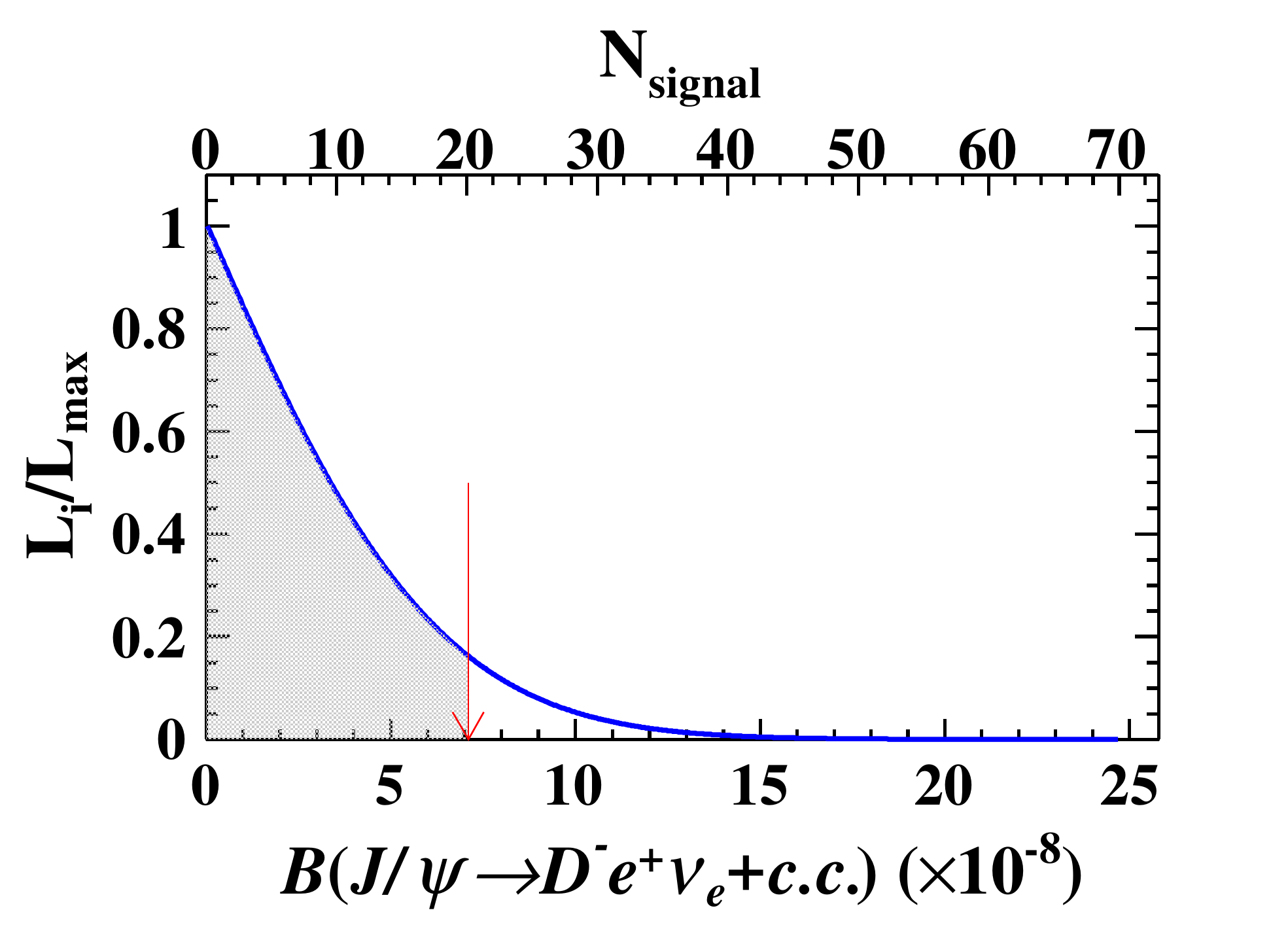}
	\caption{The distribution of the normalized smeared likelihood values (blue solid curve) as a function of the BF ($\mathcal{B}(\jpsi\to D^{-}e^{+}\nu_{e}+c.c.)$) or the number of signal events ($\rm{N_{signal}}$). 
	The shaded area corresponds to the $90\%$ CL region and the red arrow indicates the UL on the BF at 90\% CL.}
	\label{fig:smear}
\end{figure}

To set an UL on the BF via a Bayesian approach~\cite{pdg:2020, bernardo:2000}, we perform a likelihood scan with a series of fits, where the numbers of signal decays $N_{\rm{signal}}$ are fixed to values from $-$70 to 70 with a step of 0.1.
Since the BF is only meaningful in physical region ($\mathcal{B}\geq0$), the UL on the BF is calculated in this region.
To take into account any uncertainties from the choice of the fit range and the background shape of the $U_{\rm{miss}}$ distribution, we expand the fit range by $6~\mev$ on either side and simultaneously change the background shape to a second-order polynomial. The largest likelihood value is retained as the most conservative result.
Thus, we obtain the likelihood values as a function of the calculated BFs. To incorporate the systematic uncertainties described in the following section, we follow the method in Ref.~\cite{Liu:2015uha}
of combining multiple measurements of a BF, where each result can be presented as an upper limit.
%of smearing a likelihood function 
The distribution of the resulting normalized likelihood values is shown in Figure~\ref{fig:smear}.
The UL on the BF at the $90\%$ confidence level, obtained by integrating from zero to $90\%$ of the likelihood curve in the physical region ($\mathcal{B}\geq0$), is $\mathcal{B}(\jpsi\to D^{-}e^{+}\nu_{e}+c.c.)<7.1\times10^{-8}$.

%By integrating the smeared likelihood curve out to 90\% of the physical region ($\mathcal{B}\geq0$), the UL on the BF is set to be $\mathcal{B}(\jpsi\to D^{-}e^{+}\nu_{e}+c.c.)<7.1\times10^{-8}$ at 90\% CL.

%In order to get the UL on BF following the Bayesian method, a series of scan fits are performed, where the numbers of signals $N_{\rm{signal}}$ are fixed to positive values from zero to 70 with a step of 0.1.
%To consider the influences caused by the fitting range and background shape of $U_{\rm{miss}}$ on the results, we randomly change the fitting range by $\pm6~\mev$ and the background shape to a second-order polynomial, and retain the most conservative fitting result.
%Thus, the scan likelihood values as a function of the calculated BFs are obtained.
%We follow the method~\cite{Liu:2015uha} of  smearing the likelihood function to incorporate systematic uncertainties, as described in the following section, in calculating the UL on BF. The distribution of normalized smeared likelihoods is shown in Figure~\ref{fig:smear}.
%By integrating the smeared likelihood curve out to 90\% of physical region ($\mathcal{B}\geq0$), the UL on BF is set to be $\mathcal{B}(\jpsi\to D^{-}e^{+}\nu_{e}+c.c.)<7.1\times10^{-8}$ at 90\% CL.

\section{SYSTEMATIC UNCERTAINTY}
\label{sec:systematic}
\hspace{1.5em}
The main systematic uncertainties come from the tracking and PID efficiency, the signal MC model, the $E^{\rm{tot}}_{\gamma}$, $E/p$ and $|\vec{p}_{\rm{miss}}|$ requirements, the BF of the $D^-\to K^+\pi^-\pi^-$ decay and the total number of $\jpsi$ events.

\begin{itemize}
    \item \emph{Tracking and PID efficiency.} The uncertainty due to tracking and PID efficiency for kaons and pions is determined by analyzing doubly-tagged $D^{+}D^{-}$ decay events from $\psi(3770)$~\cite{bes3:kpi2017}. Using partially reconstructed hadronic decays of $D^{+}\to K^{-}\pi^{+}\pi^{+}$ and $D^{-}\to K^{+}\pi^{-}\pi^{-}$ where one $\pi^{-}$ or $K^{+}$ meson is not reconstructed, the uncertainties are estimated to be 1.0\% per track. 
In addition, the uncertainty from the positron tracking is studied using a control sample of radiative Bhabha events $\ee\to\gamma\ee$ produced at $\sqrt{s}=3.08~\gev$, while the PID uncertainty is studied using a mixed control sample of $\ee\to\gamma\ee$ events and $\jpsi\to\ee(\gamma_{FSR})$ events produced at $\sqrt{s}=3.097~\gev$.
We quote 1.0\% and 1.0\% as the systematic uncertainties on the tracking and PID efficiency for the positron, respectively.

    \item \emph{Signal MC model.} The influence of the assumed signal model on the sensitivity of the result comes from the estimation of the signal efficiency. The difference between the efficiencies estimated with the nominal model and the phase space model is taken as the systematic uncertainty, which is about 3.0\%.
    
     \item \emph{$E^{\rm{tot}}_{\gamma}$, $E/p$, and $|\vec{p}_{\rm{miss}}|$ selection requirements.} In order to estimate the systematic uncertainties due to the $E^{\rm{tot}}_{\gamma}$, $E/p$, and $|\vec{p}_{\rm{miss}}|$ selection requirements, we use a control sample of semi-leptonic signal decays $D^{0}\to K^{-}e^{+}\nu_{e}$ tagged with a $\bar{D}^{0}\to K^{+}\pi^{-}$ decay selected from $\psi(3770)$ data~\cite{Ablikim:2019hff}. We obtain the overall efficiency from a sample of 200000 signal MC simulation events. We apply the event selection criteria in Ref.~\cite{bes3:2014kenu} to the tagging mode, and the selection requirements for the positron and kaon described in Section~\ref{sec:analysis} to the signal mode.
     After applying all the requirements to the $\psi(3770)$ data sample, we get a clean signal sample with 97.8\% purity.
We perform a fit to the $U_{\rm{miss}}$ distribution to extract the signal yields and calculate the BF $\mathcal{B}(D^{0}\to K^{-}e^{+}\nu_{e})$.
By comparing the nominal result and the results without one of those requirements, we assign systematic uncertainties of $2.1\%$, $0.3\%$, and $0.3\%$ for the $E^{\rm{tot}}_{\gamma}$, $E/p$, and $|\vec{p}_{\rm{miss}}|$ requirements, respectively.

     \item \emph{BF of the $D^{-}\to K^{+}\pi^{-}\pi^{-}$ decay.} The $\mathcal{B}(D^{-}\to K^{+}\pi^{-}\pi^{-})$ result is used as an input in the baseline analysis, and its uncertainty of $1.7\%$~\cite{pdg:2020} is propagated as the systematic uncertainty.
     
     \item \emph{Number of $\jpsi$ events.} We quote a relative uncertainty of $0.5\%$ determined using $\jpsi$ inclusive hadronic decays for $N_{\jpsi}$ as the systematic uncertainty from Ref.~\cite{bes3:njpsi2012}.

\end{itemize}
All systematic uncertainties are summarized in Table~\ref{tab:syst_err}. They are added in quadrature and their total size is reported as well.
%~summarizes all of these systematic uncertainties from different sources, which are studied in the following items.
\begin{table*}[tpb]
\setlength{\abovecaptionskip}{0.0cm}
\setlength{\belowcaptionskip}{-1.6cm}
\caption{Summary of the systematic uncertainties in percentage for the measurement of the BF. The total value is calculated by summing up all sources in quadrature.}
  \begin{center}
  \footnotesize
  \newcommand{\tabincell}[2]{\begin{tabular}{@{}#1@{}}#2\end{tabular}}
  \begin{threeparttable}
  \begin{tabular}{l|c c c c c c c}
      \hline\hline
                         Sources & Relative uncertainties\\
			\hline
			Tracking & 4.0\\
			Particle ID & 4.0\\
		        Signal MC model & 3.0\\
			$E^{\rm{tot}}_{\gamma}$ requirement & 2.1\\
			$E/p$ requirement & 0.3\\
		        $|\vec{p}_{\rm{miss}}|$ requirement & 0.3\\
                         BF of the $D^{-}\to K^{+}\pi^{-}\pi^{-}$ decay & 1.7\\
			Number of $\jpsi$ events & 0.5\\
    			\hline
                         Total & 7.0\\
      \hline\hline
  \end{tabular}
  \label{tab:syst_err}
  \end{threeparttable}
  \end{center}
\end{table*}

\section{SUMMARY}
\label{sec:summary}
\hspace{1.5em}
Based upon a sample of
$10.1\times10^{9}$ $\jpsi$
events collected with the BESIII detector, the BF of the rare semi-leptonic decay $\jpsi\to D^{-}e^{+}\nu_{e}$ is studied with a semi-blind analysis.
No excess of events is observed over the background. The resulting UL on the BF is $\mathcal{B}(\jpsi\to D^{-}e^{+}\nu_{e}+c.c.)<7.1\times10^{-8}$ at $90\%$ CL, when systematic uncertainties are taken into account. Our result improves this limit~\cite{bes:2006} by a factor of 170.
This is the most sensitive search for the $\jpsi\to D^{-}e^{+}\nu_{e}$ decay.
%the $\jpsi$ rare semi-leptonic decay, 
 This measurement is compatible with the SM theoretical predictions~\cite{wang:2008b, shen:2008, dhir:2013, ivanov:2015, tian:2017}, and puts a stringent constraint on the parameter spaces for different new physics models predicting BFs of the order of $10^{-5}$~\cite{tian:2017}.

%Compared to the magnitude of $10^{-5}$ due to potential new physics contributions beyond the SM~\cite{tian:2017}, this result puts a stringent constraint on the parameter spaces for different new physics models.

\acknowledgments
\hspace{1.5em}
The BESIII collaboration thanks the staff of BEPCII and the IHEP computing center for their strong support. This work is supported in part by National Key Research and Development Program of China under Contracts Nos. 2020YFA0406400, 2020YFA0406300; Joint Large-Scale Scientific Facility Funds of the National Natural Science Foundation of China (NSFC) and Chinese Academy of Sciences (CAS) under Contracts Nos. U1932101, U1732263, U1832207; State Key Laboratory of Nuclear Physics and Technology, PKU under Grant No. NPT2020KFY04; NSFC under Contracts Nos. 11625523, 11635010, 11675275, 11735014, 11822506, 11835012, 11935015, 11935016, 11935018, 11961141012, 11975021, 12022510, 12035009, 12035013, 12061131003; the CAS Center for Excellence in Particle Physics (CCEPP); the CAS Large-Scale Scientific Facility Program; CAS Key Research Program of Frontier Sciences under Contract No. QYZDJ-SSW-SLH040; 100 Talents Program of CAS; Fundamental Research Funds for the Central Universities; INPAC and Shanghai Key Laboratory for Particle Physics and Cosmology; ERC under Contract No. 758462; European Union Horizon 2020 research and innovation programme under Contract No. Marie Sklodowska-Curie grant agreement No 894790; German Research Foundation DFG under Contracts Nos. 443159800, Collaborative Research Center CRC 1044, FOR 2359, FOR 2359, GRK 214; Istituto Nazionale di Fisica Nucleare, Italy; Ministry of Development of Turkey under Contract No. DPT2006K-120470; National Science and Technology fund; Olle Engkvist Foundation under Contract No. 200-0605; STFC (United Kingdom); The Knut and Alice Wallenberg Foundation (Sweden) under Contract No. 2016.0157; The Royal Society, UK under Contracts Nos. DH140054, DH160214; The Swedish Research Council; U. S. Department of Energy under Contracts Nos. DE-FG02-05ER41374, DE-SC-0012069.

\newpage
%CWR at 2021-xx-xx
%\author{Author list}
\begin{small}
\begin{center}
M.~Ablikim$^{1}$, M.~N.~Achasov$^{10,c}$, P.~Adlarson$^{67}$, S. ~Ahmed$^{15}$, M.~Albrecht$^{4}$, R.~Aliberti$^{28}$, A.~Amoroso$^{66A,66C}$, M.~R.~An$^{32}$, Q.~An$^{63,49}$, X.~H.~Bai$^{57}$
% ...
\\
\vspace{0.2cm}
(BESIII Collaboration)\\
\vspace{0.2cm} {\it
$^{1}$ Institute of High Energy Physics, Beijing 100049, People's Republic of China\\
% ...
$^{4}$ Bochum Ruhr-University, D-44780 Bochum, Germany\\
% ...
$^{10}$ G.I. Budker Institute of Nuclear Physics SB RAS (BINP), Novosibirsk 630090, Russia\\
% ...
$^{15}$ Helmholtz Institute Mainz, Staudinger Weg 18, D-55099 Mainz, Germany\\
% ...
$^{28}$ Johannes Gutenberg University of Mainz, Johann-Joachim-Becher-Weg 45, D-55099 Mainz, Germany\\
% ...
$^{32}$ Liaoning Normal University, Dalian 116029, People's Republic of China\\
% ...
$^{49}$ State Key Laboratory of Particle Detection and Electronics, Beijing 100049, Hefei 230026, People's Republic of China\\
% ...
$^{57}$ University of Jinan, Jinan 250022, People's Republic of China\\
% ...
$^{63}$ University of Science and Technology of China, Hefei 230026, People's Republic of China\\
% ...
$^{66}$ University of Turin and INFN, (A)University of Turin, I-10125, Turin, Italy; (B)University of Eastern Piedmont, I-15121, Alessandria, Italy; (C)INFN, I-10125, Turin, Italy\\
$^{67}$ Uppsala University, Box 516, SE-75120 Uppsala, Sweden\\
% ...
$^{c}$ Also at the Novosibirsk State University, Novosibirsk, 630090, Russia\\
% ...
}\end{center}

\vspace{0.4cm}
\end{small}
\newpage

%\end{linenumbers}
\end{document}